\documentstyle[12pt]{article}

\newcommand{\be}{\begin{equation}}
\newcommand{\ee}{\end{equation}}
\newcommand{\bea}{\begin{eqnarray}}
\newcommand{\eea}{\end{eqnarray}}

\newcommand{\s}{\sigma}

\begin{document}

\begin{titlepage}
\begin{center}
\vskip .2in
\hfill
\vbox{
    \halign{#\hfil         \cr
           hep-th/9805041 \cr
           SU-ITP-98-33 \cr
           SLAC-PUB-7820\cr
           May  7 1998   \cr
           }  
      }   
\vskip 2cm
{\large \bf VACUA OF M-THEORY  AND STRING THEORY}\\
\vskip 1.0cm
{\bf Renata Kallosh\footnote{email address: kallosh@physics.stanford.edu}
and
Arvind Rajaraman}\footnote{email address: arvindra@leland.stanford.edu}
\footnote{Supported in part by the Department of Energy under contract 
No. DE-AC03-76SF00515.}

\vskip 1.0cm

{\em Department of Physics, Stanford University, Stanford, California
94305
and
\\}
\vskip 0.5cm

{\em Stanford Linear Accelerator Center, Stanford University, Stanford, \\
CA
94309 USA \\}
\vskip 1cm

\end{center}
\begin{abstract}
We argue that supersymmetric higher-dimension operators in the effective
actions
of M-theory and IIB string theory do not affect the maximally
supersymmetric
vacua: $adS_4\times S^7$ and $adS_7\times S^4$ in M-theory and
$adS_5\times
S^5$
in IIB string theory. All these vacua are described in superspace by a
fixed point with all components of supertorsion and supercurvature being
supercovariantly constant. This follows from 32 unbroken
supersymmetries and allows us to prove that such vacua are exact.

\vskip 0.5cm
 \end{abstract}
\end{titlepage}
\newpage

\section{Introduction}

There is a very limited knowledge of exact solutions in gravitational
theories
which include higher-dimension operators. An example of such a
configuration
is
given by a pp-wave. This solves the non-linear equations of motion of pure
Einstein theory and can be proved to remain an exact solution in the
presence
of
all possible higher-derivative terms respecting general covariance.

It is interesting to find some  solutions in M-theory and string theory
which
can be proved to be exact when all possible corrections to the low-energy
supergravity actions are included, which respect not only the general
covariance but also the local supersymmetry. It is natural to consider the
vacuum solutions and use the power of 32 unbroken supersymmetries.

We shall look at $adS_4\times S^7$ and $adS_7 \times S^4$ solutions of
M-theory
and $adS_5\times S^5$ solution of IIB string theory. There has been a
great
deal
of interest in these solutions lately because of the conjecture
\cite{juan,polyakov,witten} relating IIB string theory on $adS_5\times
S^5$ to
N=4 Yang Mills theory. We shall attempt to argue that there are no
corrections
to the form of this solution from $\alpha'$ corrections. This was already
shown
for the $adS_5\times S^5$ case to order $\alpha'^{3}$ in
\cite{banksgreen}.
Similarly, we argue that there are no $l_{11}$ corrections to the form of
$adS_4\times S^7$ and $adS_7\times S^4$ in M-theory. The proof in
\cite{banksgreen} uses essentially the conformal flatness of $adS_5\times
S^5$
space. Our general proof based on the maximal amount of unbroken
supersymmetry
will cover the supersymmetric vacua of M-theory whose metrics are not
conformally flat.

In the case of the M-theory solutions, we still do not have a full
formulation
of the theory. However, we can study the low energy effective action as an
expansion in powers of the Planck length. We expect that the  effective
action
will have $N=1$ supersymmetry  in eleven dimensions, which constrains its
form.
Also, in analogy with string theory, we expect that an exact solution of
the
effective action is a solution of the full theory.

The strategy will be to write down all possible corrections to the
equations
of
motion consistent with supersymmetry. If all possible corrections to the
equations vanish when evaluated in a certain background, then, by
definition,
this background is an exact solution of the full effective action. We
shall
show
that this situation holds for these solutions.

In general, the  possible corrections to the equations of motion could
involve
the curvature, derivatives of the curvature etc. It is a feature of these
solutions that all relevant tensors are covariantly constant. Hence the
corrections can only depend on the numerical values of these tensors.
We will then show that even these corrections do not affect the solutions.

This is most conveniently done in superspace. We find that in superspace,
the
equations of motion can be written in a form which have one free spinorial
index.
It turns out that all nonzero components of the superfields (in this
background) have two spinor indices, and it is thus impossible to
construct a consistent nonzero correction.
(Usually of course, one could have used spinorial derivatives to construct
a
correction term, but as we have already said, all such terms vanish.) Thus
the
solution is uncorrected in the full effective action.

We first consider as a warm-up, the cases of pp-waves in pure gravity, and
the
$adS_2\times S^2$ solution of  $N=2, d=4$ pure supergravity, where similar
considerations allow us to prove the exactness of the solutions. We then
turn
to
the cases of interest i.e. $adS_4\times S^7$ and $adS_7\times S^4$ in
eleven-dimensional supergravity and $adS_5\times S^5$ in IIB string
theory.
Finally we conclude with discussions.

Recently, quantum corrections to the supersymmetric black hole entropy in
string
theory \cite{MSW} and to the minimal value of the central charge in
supergravity theory \cite{Klaus} have been calculated. These corrections
appear
in theories related to $N=2$ supergravity interacting with vector
multiplets.
Such  interaction is not unique. The prepotential in presence of higher
dimension
operators is modified \cite{Klaus} but the theory is still
supersymmetric. It
has not been established whether the existence of such corrections is due
to
the
modifications of the solutions or just change of the $adS_2$ size  in the
Bertotti-Robinson throat. In all cases which we will study we will deal
only
with maximal supersymmetry, 32 in d=11, d=10 (and 8 in $d=4$ for pure
supergravity
without extra matter multiplets as a simplest model). These are purely
geometric
theories in superspace. There are no options in the choice of the
prepotential.
We expect therefore that the superfield structure is not modified in
presence
of
corrections.

\section{Stability  of  pp  waves }

Pp-wave geometries are space-times admitting a covariantly
constant null vector field
\begin{equation}\label{null}
\nabla_{\mu} l_{\nu} = 0\ ,  \qquad l^{\nu}l_{\nu}= 0 \ .
\end{equation}
Spacetimes with this property were first discovered by Brinkmann
\cite{Br1}.
The existence of a covariantly constant null vector field has
dramatic
consequences \cite{Gu1}.
For instance, for the class of d-dimensional pp-waves
with metrics
of the form  \begin{equation}\label{d}
ds^2 = 2 du dv + K (u, x^i ) du^2 - dx^i d x^i \ ,
\end{equation}
where $i, j = 1,2, ..., d-2$, the Riemann curvature is \cite{Gu1}
\begin{equation}
R_{\mu\nu\rho\sigma} = - 2 l_{[\mu}( \partial_{\nu]}
\partial_{[\rho} K )
l_{\sigma]}\ .
\end{equation}
The Ricci tensor vanishes if $K$ is a harmonic
function in the transverse space:
\begin{equation}
R_{\mu \sigma} = - {1\over 2} ( \partial_{\nu}
\partial^{\nu} K )
 l_{\mu}l_{\sigma}\ , \qquad R= - {1\over 2} ( \partial_{\nu}
\partial^{\nu} K )
 l_{\mu}l^{\mu}=0
\end{equation}
The curvature $R_{\mu\nu\rho\sigma}$ is therefore orthogonal to $l^{\mu}$
and
to
$\nabla^
\mu $ in all its indices.
Since $K$ is independent of $v$, the metric solves Einstein equations
$G_{\mu\nu}=0$ if  $\partial^2_T K=0$. Possible corrections to field
equations
may come from
higher dimension operators and depend on the curvature tensors and their
covariant derivatives
\begin{equation}
G_{\mu\nu}= F_{\mu\nu}^{corr} ( R_{\mu\nu\lambda\sigma},
D_{\alpha} R_{\mu\nu\lambda\sigma}, \dots)
\end{equation}
 Corrections to Einstein equations are quadratic or higher order in
curvature
tensors. However, there is no way to contract two or more of Riemann
tensors
which will form a two-component tensor to provide the r.h.s. of the
Einstein
equation coming from higher dimensions operators. Therefore all higher
order
corrections vanish for pp-waves solutions. They remain exact solutions of
any
higher order in derivatives general covariant theory. This includes
supergravities and string theory with all possible sigma model and string
loop
corrections to the effective action, as long as these corrections respect
general covariance. Note that supersymmetry played no role in establishing
this
non-renormalization theorem.

\section{Supersymmetric Bertotti-Robinson vacuum}

Our next example is N=2, d=4 pure supergravity without matter
multiplets. A
vacuum solution with 8 unbroken supersymmetries is given by the $adS_2
\times
S^2$ metric and a two-form which is a volume form of the $adS_2$
space. Before
considering N=2 theory we will explain our strategy in terms of the more
familiar superspace of N=1 supergravity.

In general the geometric superspace tensors must satisfy  some constraints
in
order to describe correctly the field contents of supergravity
theory. When
the
constraints are imposed, the geometric Bianchi identities are not
identities
anymore but equations which can be solved. The solutions provide the
superspace
form of supergravities. In N=2 d=4 case the full off-shell superspace
solution
is available. This is  analogous to the well known N=1 d=4 supergravity in
superspace given in terms of 3 superfields: $W_{\alpha \beta \gamma},
G_{\alpha
\dot \beta}, {\cal R }$. All components of the constrained geometric
tensors
like torsion $T_{AB}^C$ and curvature $R_{AB}{}^{cd}$ are expressed in
terms
of
these 3 superfields and their covariant derivatives. On shell $G_{\alpha
\dot
\beta}(X, \theta)=0$ and also ${\cal R }(X, \theta) =0$. All possible
higher
dimension operators would modify the form of classical equations of motion
as
follows
\begin{equation}
G_{\alpha \dot \beta} = {\cal F}^{corr}_{\alpha \dot \beta} (G , {\cal R}
, W,
\bar W , D_A G, D_A R, D_A W, D_A \bar W, \dots  ),
\end{equation}
It is expected that the RHS of the quantum corrected equation of motion
will
depend
only
on superfields and their covariant derivatives, i.e. on all supertensors
of
the
theory. If one wishes to find out if some particular solution of
classical equations remains a solution in the presence of the corrections,
one
has to study whether
\begin{equation}
 {\cal F}^{corr}_{\alpha \dot \beta} (G=0 , {\cal R}=0 , W, \bar W , D_A
G=0,
D_A R=0, D_A W, D_A \bar W, \dots  ),
\end{equation}
vanishes or not. The chiral superfield $W_{\alpha \beta \gamma}$ has in
the
lowest component $\theta^0$ the gravitino field strength and in the first
one
$\theta^1$  the Weyl tensor.

 We  proceed to N=2 d=4 case to study the supersymmetric Bertotti-Robinson
vacuum.
We give below a summary on N=2 d=4 off shell superspace with 4 bosonic and
8
fermionic
coordinates. The supergeometry
is given in \cite{gates} and we use the  two-component spinor notation
from
there. The structure group consists of  a Lorentz transformations with
$M_{ab}=
-M_{ba}, a=0,1,2,3$ and central charge transformations $M_{ij}=-M_{ji},
i,j=1,2$. The geometric tensors include torsion $T^A_{BC}$, the Lorentz
curvature $R_{ab}{}^{cd}$ and the central charge curvature
$F_{AB}{}^{ij}$.

There are 2  superfields defining the off-shell superspace. There is one
spinorial
superfield $T_{\alpha}^i (X,\theta,  \bar \theta)$ which vanishes on-shell
and
therefore represents the
superfield equations of motion of the theory. There is also a chiral
superfield
$W_{\alpha \beta ij }$ satisfying $D_{\dot \gamma k} W_{\alpha \beta ij
}=0$.
The lowest $\theta^0$ component of the superfield $W$ is the form field,
the
next one  $\theta^1$   is the gravitino field strength and the second one
$\theta^2$  is the Weyl tensor
\begin{eqnarray}
W_{\alpha \beta\; ij} (X, \theta)|_{\theta=0} &=& (\sigma ^{ab})_{\alpha
\beta}
F_{ab\; ij}(X)\\
\nonumber\\
D_{ \alpha }^i W_{ \beta \gamma \; i k }(X, \theta)|_{\theta=0} &=&
\psi_{\alpha \beta \gamma \; k}(X)\\
\nonumber\\
D_{ \alpha }^i D_{\beta }^j W_{ \gamma \delta \; i j }(X,
\theta)|_{\theta=0}
&=&  C_{\alpha \beta \gamma \delta} (X)\\
\nonumber\\
D_{ \dot \alpha }^i D_{\beta  \; i} W_{ \gamma \delta \; kl }(X,
\theta)|_{\theta=0} &=&  D_{\dot \alpha \beta} F_{\gamma \delta \; kl }
(X)
\end{eqnarray}

According to our conditions on the corrections to field equations
respecting N=2 supersymmetry we get the quantum corrected field equation
in
the
form
\begin{equation}
T_{\alpha}^i (X, \theta) = {\cal F}^{i \; corr}_{\alpha} (T ,  W, \bar W ,
D_A
T , D_A W, D_A \bar W, \dots  ),
\end{equation}

{\it Exactness of flat superspace}. Flat superspace has the following
properties. There is a non-vanishing constant torsion and central charge
curvature.
\begin{eqnarray}
T_{\alpha i, \dot \beta j}^{d} = 2i \sigma_{\alpha \dot \beta}
\delta_{ij}\ ,
 \qquad  F_{\alpha i, \beta j }{}^{kl} = C_{\alpha \beta} \delta_ i^{[k}
\delta_j^{l]} \ .
 \label{flat} \end{eqnarray}
The superfields   $T_{\alpha}^i (X, \theta) \ , W_{ab}^{ij}(X, \theta) $
vanish. If one would try to construct ${\cal F}^{i \; corr}_{\alpha}$ out
of
only constant structures in eq. (\ref{flat}), one could see that no such
structures are available and therefore the flat superspace can not have
quantum
corrections.

A superspace form of the near horizon  black hole geometry with a 2-form
and
with enhancement of supersymmetry near the horizon has been studied before
\cite{K, FK}. It has been found that the supersymmetric BR vacuum
corresponds
to  a supercovariantly constant superfield $W_{ab}$ (the  superfield
$T_{\alpha}^i (X,\theta,  \bar \theta)=0$ since we consider the solution
of
the
classical field
equations)
\begin{equation}
 D_A W^{BR}_{ab\; kl} =0\quad  \Longrightarrow  \quad D_{\alpha i}
W^{BR}_{ab
\; kl} =  D_{\dot \alpha i} W^{BR}_{ab \; kl} =
D_{c} W^{BR}_{ab \; kl}=0
\label{const}\end{equation}
The integrability condition for the existence of the covariantly constant
superfield is verified by checking that the solution admits Killing
spinors of
the maximal dimension. It can also be simply understood by observing that
for
the supersymmetric BR the lowest  $\theta^0$ component of the superfield
is
covariantly constant in $X$-space, the next $\theta^1$ component vanishes
since
the background is bosonic and the second  $\theta^2$ component of the
superfield vanishes since the Weyl tensor vanishes and the form $F$ is
covariantly constant in $X$-space. The higher components of the superfield
are
not independent and therefore also vanish. The self-dual form is
\begin{equation}
F^{ij} = \epsilon^{ij} ( e^0 \wedge e^1 + e^2\wedge e^3)
\end{equation}
Therefore all components of the superfield $W$ vanish except the first one
which is a constant self-dual form. It breaks the Lorentz part of the
structure
group $SO(1,3)$ of the superspace with $a=0,1,2,3$ into a product
$SO(1,1)\times SO(2)$, with $\hat{a} =0,1$ and $\check{a}=3,4$. The first
one
is related to the tangent space of $adS_2$ and the second one to that of
$S^2$.

Thus our {\it BR vacuum in the superspace} can be described by a
covariantly
constant superfield $W_{ {a} {b}}^{BR}$ which consist of 2 parts:
\begin{equation}
W_{\hat {a} \hat {b}}^{BR}  =  \epsilon_{\hat {a} \hat {b}}\ , \qquad
W_{\check{a} \check{b}   }^{BR} = \epsilon_ {\check{a} \check{b}}
\label{BR}\end{equation}
All non-vanishing components of torsion and curvature are constant and
given
by
eq. (\ref{flat})
as in the flat superspace as well as new constant torsions and curvatures:
\begin{eqnarray}
&&    T_{ a, \beta j, \dot \gamma k} = -i \sigma ^{ b}_{\beta , \dot
\gamma }
W_{{a}  {b}}^{BR}\ ,  \qquad F_{ {a} {b}  }^{kl} = \epsilon^{kl}W_{ {a}
{b}}^{BR} \ , \\
\nonumber\\
&&R_{\alpha i, \beta j }{}^{cd} = - 2i C_{\alpha \beta} (\bar
{\sigma}^{cd}
\bar W^{BR})_{\dot \delta} ^{\dot \delta} \ , \qquad {\rm etc.}
\end{eqnarray}
Now we can look what will happen with corrections to the equation of
motion with
account of (\ref{const}) and (\ref{BR}).
\begin{equation}
T_{\alpha}^i (X, \theta) = {\cal F}^{i \; corr}_{\alpha} (  W_{\hat {a}
\hat
{b}}^{BR}  , W_{\check{a} \check{b}   }^{BR}   ),
\end{equation}
It is not possible to build the object ${\cal F}^{i \; corr}_{\alpha}$
with
one
fermionic index from the available supercovariantly constant superfields.
Therefore
we do not see any possibility for  the supersymmetric BR vacuum to be
corrected
by higher dimension supersymmetric operators.

\section{ $adS_4\times S^7$ and $adS_7\times S^4$ vacua of M-theory}

The background is in the  $AdS_4$ case
\bea
F^{(AdS)}_{mnps}=e\epsilon_{mnps}\\
R^{(AdS)~~ps}_{~~~~~mn}=-{4e^2\over
9}(\eta_m^p\eta_n^s-\eta_m^s\eta_n^p)\\
R^{(Sph)~~ps}_{~~~~~mn}={e^2\over 9}(\eta_m^p\eta_n^s-\eta_m^s\eta_n^p)
\eea
and for the $AdS_7$ case
\bea
F^{(Sph)}_{mnps}=e\epsilon_{mnps}\\
R^{(AdS)~~ps}_{~~~~~mn}=-{e^2\over 9}(\eta_m^p\eta_n^s-\eta_m^s\eta_n^p)\\
R^{(Sph)~~ps}_{~~~~~mn}={4e^2\over 9}(\eta_m^p\eta_n^s-\eta_m^s\eta_n^p)
\eea

The relevant on-shell superspace was constructed in \cite{CremFer,
BrinkHowe}.
There
is a single
superfield
$W_{rstu}(X, \theta) $\footnote{We follow the notation of \cite{CremFer}
with the
exception of renaming spinorial indices in tangent space from $a$ to
$\alpha$
to be in agreement with other sections of this paper.}.
The field content of this superfield follows from that
of
eleven-dimensional
supergravity.

The first few  components of the superfield are
\begin{eqnarray}\label{0}
W_{rstu} (X, \theta)|_{\theta=0} &=& F_{rstu}
(X)\\
\nonumber\\
\left ( D_{ \alpha } W_{rstu  }(X, \theta)\right ) |_{\theta=0} &=& 6
(\gamma_{[rs}
\hat D_t  \psi_{s})_\alpha (X)\label{1}
\\
\left (D_{ \alpha } (\hat D_{[r } \psi_{s]})_\beta\right )|_{\theta=0}
&=& (
{1\over
8}
\hat R_{rsmn} (X)  \gamma^{mn}+
{1\over 2} [T_r^{tuvw}, T_s^{xyzp}] \hat
F_{tuvw} (X) \hat F_{xyzp}
\nonumber\\
& & ~~~~~~~~~~~~~~+ T_{[s}^{tuvw} \hat D_{r]} \hat F_{tuvw}(X)
)_{\alpha\beta}
\label{2}
\end{eqnarray}
Here  $T^{rstuv}$ is a product of $\gamma$-matrices defined in
\cite{CremFer}.

The  equation of motion of classical supergravity in superspace is
\bea
(\gamma^{rst}D)_\alpha W_{rstu}(X,\theta)=0
\eea

In a generic background one can write down corrections to the RHS of the
superfield equations
involving the superfields, derivatives of the superfield etc.
There is no reason to expect that such corrections will vanish in general.

We now claim that the supersymmetric $adS_4\times S^7$ and $adS_7\times
S^4$
vacua of M-theory are described by a fixed point in superspace, where all
components of torsion, curvature and 4-form are covariantly constant. To
prove
this it is sufficient to prove that the superfield $W_{rstu} (X,\theta)$
is
supercovariantly constant (since all other superfields can be derived from
it.)

The lowest component
of the superfield $W$ according to eq. (\ref{0}), is given by the form
field strength. In the
$AdS_7$ case, we have $F_{0123}=\epsilon _{0123}$, and in the
$AdS_4$ case, we have $F_{45678910}=\epsilon_{45678910}$. These
are manifestly covariantly constant.

The next component of the superfield, as shown
in eq. (\ref{1}), is the gravitino field strength and  this vanishes since
our
vacua
are purely bosonic.

The next component of the superfield is bosonic and  is
shown in eq. (\ref{2}). Remarkably, it vanishes as well (as can be
verified
by explicit computation.)

The remaining higher components are given by some derivatives of
the
previous ones and therefore all vanish. Putting these facts together, we
see
that
the superfield $W_{rstu} (X,\theta)$ is supercovariantly constant.

 The vanishing of the $\theta^2$ component of the superfield   is related
to the fact that these vacua have maximal supersymmetry. The
integrability
condition for the requirement that the bosonic
configuration admits maximal unbroken 32-dimensional supersymmetry is
\begin{equation}
\delta_{SUSY} \psi_r = D_r \epsilon + T_r^{tuvw} \epsilon F_{tuvw}=0
\end{equation}
It was shown in  \cite{KK} (in the context of the study of the near
horizon
Killing
spinors of M2 and M5 branes) that this equation yields
\begin{eqnarray}\delta_{SUSY}  (\hat D_{[r } \psi_{s]})  &=& {1\over
8}
\hat R_{rsmn}  \gamma^{mn} \epsilon + {1\over 2} [T_r^{tuvw}, T_s^{xyzp}]
\epsilon  F_{tuvw}
  F_{xyzp} \nonumber\\
&+& T_{[s}^{tuvw} \hat D_{r]} \epsilon   F_{tuvw}  )=0\label{enh}
\end{eqnarray}
which is exactly the statement that the $\theta^2$ component vanishes.

Thus we have shown that the
integrability condition for the 32 Killing spinors of the vacua provides
the
proof that the superfield is covariantly constant.
\begin{equation}
 D_A W_{rstu} =0\quad  \Longrightarrow  \quad D_{a} W_{rstu} =
D_{v} W_{rstu}=0
\label{constM}\end{equation}

Let us look now at the corrected equations of motion. Since
$D_A
W_{rstu}=0$ the corrections can depend only on $W_{rstu}$ and other
constant
tensors like $\gamma$-matrices etc. Again we observe that it is impossible
to
get one spinorial index
without using spinorial
derivatives, but  such derivatives are zero on all the terms. Hence there
is no possible correction we can write down. This shows that the
$adS_4\times
S^7$ and $adS_7 \times S^4$ solutions are exact.

\section{$adS_5\times S^5$ vacuum of string theory}

We have, in this case, to consider the superspace formulation of
type IIB supergravity. This was constructed in \cite{HoweWest}.

The background has a nonzero five-form field strength and a nonzero
curvature.
These split into the AdS part and the sphere part. For the AdS part, we
have
\bea
g^{(AdS)}_{mnpst}=e\epsilon_{mnpst}\\
R^{(AdS)~~ps}_{~~~~~mn}=-{e^2\over 16}(\eta_m^p\eta_n^s-\eta_m^s\eta_n^p)
\eea
where the indices run over the AdS indices (0 to 4), and for the sphere
part,
we
have
\bea
g^{(Sph)}_{mnpst}=e\epsilon_{mnpst}\\
R^{(Sph)~~ps}_{~~~~~mn}={e^2\over 16}(\eta_m^p\eta_n^s-\eta_m^s\eta_n^p)
\eea
where the indices now run over the sphere indices (5 to 9.)
The important point about these values is that again, all the tensors are
covariantly
constant in $X$-space.

The  on-shell superspace description of  IIB string theory is related to
$N=2,
d=10$ chiral supergravity \cite{HoweWest}. The superspace has some
constrained
torsion $T_{AB}^C$, , Lorentz curvature $R_{AB}^{cd}$
and $U(1)$ curvature $M_{AB}$. Besides, there are the 3-form $F_{ABC}$,
the
5-form $G_{ABCD}$ and the scalar field strength $P_A$.

In the full non-linear theory
there are two  superfields,  $\Lambda_\alpha (X,\theta, \bar \theta)$ and
$Z^{+}_{abcde} (X,\theta, \bar \theta)$.  All geometric tensors are
functionals of these superfields and their covariant derivatives.
$\Lambda_\alpha
(X,\theta, \bar \theta )$ starts with the dilatino and $Z^{+}_{abcde} =
{1\over 192} G_{abcde} $ starts with the self-dual 5-form ${1\over 192}
g_{abcde}(x)$.
Even though there is only one supermultiplet, the second superfield is not
a
derivative of the first.
The scalars of this theory belong to the coset space
of ${SU(1,1) \over U(1)}$. The construction in fact starts with the
superfield
$V(X,\theta, \bar \theta)$ which is an element of  $SU(1,1)$. From this a
$SU(1,1)$ singlet  $P_A$
is built where the scalars appear. In this form scalars can be found in
derivatives of $\Lambda_\alpha (X,\theta, \bar \theta)$.

In the linear approximation one can also consider an analytic superfield
$A$
with $\bar D_\alpha A=0$ and the constraint $D^4 A = \bar D^4 \bar
A$. This
superfield in the proper basis depends only on half of the components of
the
superspace. The superinvariants of the type $R^4$ can be analysed as
superspace
integrals over 16 $\theta$. The  $\theta^4$ component of this linear
superfield
is
a Weyl tensor. This automatically proves that the higher dimension
operator
with 4 powers of the Weyl tensor will not change the background, which is
conformally invariant \cite{banksgreen}.
In what follows we will not use the linearized approximation and study the
full
theory.

The first step, as before, is to prove that all the superfields are
supercovariantly
constant in this background.
For the superfield $\Lambda_\alpha
(X,\theta, \bar \theta )$, the lowest component is the dilatino, which
automatically
vanishes in this background. The next component involves the three-form
field
strength,
which is also automatically zero. The following component is the gravitino
field
strength which is also zero. However, at order $\theta^3$ in the
superfield,
we
have a non-trivial expression involving the curvature. We must
show that this expression is zero.

The story is similar for the second superfield $Z^+_{abcde}$. The lowest
(bosonic) component is the five-form field strength, which, as mentioned
before,
is covariantly constant in our vacuum. The next component is the gravitino
field strength,
which
vanishes. However, at order $\theta^2$, we obtain a non-trivial expression
involving
the curvature. Again, we must show that this expression is zero.

It is also sufficient to prove that these two problematic expressions
vanish.
All higher components of these superfields are related to derivatives
of the components already referred to. Hence, if we can show that these
problematic
expressions vanish, we will have shown that the superfield
$\Lambda_\alpha$
is identically zero, and that the superfield $Z^+_{abcde}$ is
supercovariantly constant.

Actually, since both these problematic expressions are preceded in the
superfield by the gravitino field strength, they are related to each other
and to the variation of the gravitino field strength under supersymmetry
transformations.

We will again use the existence of maximal supersymmetry in this
background to
help us analyze this situation.
The Killing spinor equation is
\begin{equation}
\delta_{SUSY} \psi_r = \nabla_r \epsilon -i {1\over 192}
g_{rabcd}\sigma^{abcd} \epsilon =0
\end{equation}
As in the previous case of M2 and M5 branes near the horizon, the
integrability
condition for the existence of such 32 spinors for the D3 branes near the
horizon was established in \cite{KK}. This transfers to the statement that
for
the
supersymmetric $adS_5\times S^5$ vacuum we have
\begin{eqnarray}
\delta_{SUSY}   \hat \nabla_{[r } \psi_{s]}  =0
\end{eqnarray}
This is the integrability condition for the requirement that the bosonic
configuration admits maximal unbroken 32-dimensional supersymmetry.

What we see is that the variation of the gravitino field strength
vanishes.
This
also implies that the problematic expressions in the two superfields also
vanish.
This then implies that the superfields are supercovariantly constant.

Let us look at this from the superspace perspective. The gravitino field
strength forms a $T_{ab}^\delta$ component of the torsion tensor and
$T_{b\gamma}^\delta$ is a function of the form field. The superspace
Bianchi
identity defines the fermionic derivative of the torsion
through
\bea
R_{ab,\gamma}{}^\delta &=& D_{\gamma} T_{ab}^\delta +\{D_a
T_{b\gamma}^\delta
+\nonumber\\ &&T_{a\gamma}^{~~\epsilon}T_{b\epsilon}^{~~{\delta}}-
T_{a\gamma}^{~~\bar{\epsilon}}T_{b\bar{\epsilon}}^{~~{\delta}}
 - (a-b)\}- i
\delta_\gamma^\delta M_{ab}={1\over 4} (\sigma^{cd})_\gamma^\delta
R_{ab,cd}
\eea
The term $D_{\gamma} T_{ab}^\delta$ vanishes due to the Killing spinor
equation,
the term $D_a T_{b\gamma}^\delta$ vanishes since our form is covariantly
constant in $X$-space. Finally $M_{ab}$ vanishes for our background. We
are
left
with
\bea
R_{ab,\gamma}{}^\delta =
T_{a\gamma}^{~~\epsilon}T_{b\epsilon}^{~~{\delta}}-
T_{a\gamma}^{~~\bar{\epsilon}}T_{b\bar{\epsilon}}^{~~{\delta}}
 - (a-b)\}={1\over 4} (\sigma^{cd})_\gamma^\delta R_{ab,cd}\label{integra}
\eea
This coincides with the integrability condition for the existence of 32
unbroken supersymmetries and proves that the superfield $Z^{+}_{abcde}
(X,\theta, \bar
\theta)$ is covariantly constant and that all components of the superfield
$\Lambda_\alpha (X,\theta,
\bar \theta)$ vanish.

To prove that the  $adS_5\times S^5$ vacuum is exact we have to study the
possibilities
to modify the equations of motion in this vacuum.

The equation of motion are those for the dilatino superfield and the one
for
the gravitino as in previous cases. The equation of motion for bosonic
fields
come out as some higher components of these fermionic equations. Following
the
same reasoning as in previous cases we may conclude that higher dimension
supersymmetric
operators can not modify this vacuum defined by a covariantly constant
superfield.

\section{New Supergeometries}
In this section, we will present a description of the $AdS_7 \times S_4$,
$AdS_4 \times
S_7$
and $AdS_5 \times S_5$ geometries in superspace. This provides an
invariant
description of these geometries, much as the
equation $R_{rstu}=-k^2(\eta_{rt}\eta_{su}-\eta_{ru}\eta_{st})$ provides
an invariant description of anti-de-Sitter geometry.

We begin with the two M-theory solutions.

In the coordinate system in which the lowest component is also
independent of $X$ the superfield is given by a constant completely
antisymmetric tensor, for p=2
\begin{equation}
 W^{el. vac}_{\hat r \hat s \hat t \hat u} =\epsilon_ {\hat r \hat s \hat
t
\hat u}   , \qquad \hat r , \hat s = 0,1,2,3. \label{4form}\end{equation}
and for p=5 by a dual one  
\begin{equation}
 W^{ma. vac}_{\hat r \hat s \hat t \hat u} =
i \epsilon_{\hat r \hat s \hat t \hat u}   \label{dualform}\end{equation}

These tensors break the structure group of the superspace $SO(1,10)$ to
the
product $SO(1,3)\times SO(7)$ and $SO(1,6)\times SO(4)$, respectively.
Now we can give a superspace definition of the $adS_4\times S^7$ and
$adS_7\times S^4$ vacua of M-theory where all components of torsion,
curvature
and forms are covariantly constant. In addition to the flat superspace
structures, which are independent on $W$,
 we have
few more {\it $X, \theta$-independent components of  supercurvature and
supertorsion}
(we only give the nonzero values)
\bea
T^r_{\alpha\beta}=-{i\over 2}(\gamma^0\gamma^r)_{\alpha\beta} \ , \qquad
F_{rs\alpha\beta}= -{1\over 2}
(\gamma^0\gamma_{rs})_{\alpha\beta} \\
T^\gamma_{\alpha r}={1\over 2}W^{vac}_{pstu}(T_r^{pstu})^\gamma_\alpha\ ,
\quad\quad
R_{\alpha\beta}^{mn}=(\gamma^0S)_{\alpha\beta}^{mnuvzw}W^{vac}_{uvzw} \\
\nonumber\\
R_{rs}{}^{\beta\gamma} = {1\over 4} R_{rs}^{mn}
(\gamma_{mn})^{\beta\gamma} = -
[T_r^{tuvw},
T_s^{xyzp}] W^{vac}_{tuvw}   W^{vac}_{xyzp} \label{curv}
\eea
where for the constant tensors $W^{vac}_{rstu}$ we have to substitute
their
values (\ref{4form}) or (\ref{dualform}) for each vacuum.
The value of the space-time curvature in eq. (\ref{curv}) precisely shows
that
the Killing spinor integrability equation (\ref{enh}) is satisfied since
$D_r
F_{tuvw}=0$ for both  vacua.

For the $AdS_5 \times S_5$ background, we have
\bea
T_{\alpha \bar \beta}^c = -i (\sigma^c)_{\alpha \beta} \qquad F_{a\beta
\gamma} = -i (\sigma_a)_{ \beta \gamma} \\
 F_{a\bar \beta \bar \gamma} = -i (\sigma_a)_{ \beta \gamma} \qquad G_{abc
\alpha \beta} = (\sigma_{abc})_{\alpha  \beta } \\
T_{a\beta}^\gamma={i\over 192}(\s^{bcde})_\beta^\gamma g_{abcde}\\
T_{a\bar{\beta}}^{\bar{\gamma}}={i\over 192}(\bar{\sigma
}^{bcde})_\beta^\gamma
 g_{abcde}\\ R_{\alpha\bar{\beta},ab}=-{1\over
24}(\sigma^{cde})_{\alpha\beta}g_{abcde}\\ R_{ab,\gamma}{}^\delta =
T_{a\gamma}^{~~\epsilon}T_{b\epsilon}^{~~{\delta}}-
T_{a\gamma}^{~~\bar{\epsilon}}T_{b\bar{\epsilon}}^{~~{\delta}}
 - (a-b)\}
\eea

\section{Discussion}

We have established that the $adS_{p+2}\times
S^{d-p-2}$ vacua of M-theory and string theory are uncorrected by
higher-dimension supersymmetric operators. Thus we have 3 distinct vacua
in
M-theory, flat superspace, that of the near horizon M2 brane and that of
the
near horizon M5 brane. In string case we have 2 vacua, the flat superspace
and
that of the near horizon D3 brane\footnote{It has been anticipated in
\cite{TseyMets} that the exactness of $adS_{5}\times
S^{5}$  may be derived using 32 unbroken supersymmetries.}. The  $X$ space
geometry of these configurations, $adS_{p+2}\times
S^{d-p-2}$ with forms was found in \cite{GT}. Here we found the
supergeometry
of these 3 vacua of M-theory and 2 vacua of string theory. Since all the
components
of torsion and curvature in superspace for all these vacua  are found to
be
supercovariantly constant (and actually constant in the coordinate system
related
to the near horizon geometry of branes) we concluded that there are no
corrections
modifying such vacua.

 Although we have established that the form of the geometry is
unchanged, we cannot {\it a priori} exclude a change in the values of the
parameters. We believe, however, that in these cases, the Dirac
quantization
condition fixes  the flux of the field strength through the
sphere to be an integer, and thus the flux should not be affected
by small deformations. This fixes the parameters of the solution in terms
of the Planck length. In addition, the Planck length may itself be
renormalized
from its bare value, because we cannot exclude, via this analysis,  the
appearance
in the effective action of terms proportional to the original equations of
motion (which vanish on-shell).

In {\it even dimensions for the self-dual vacua  $adS_{5}\times
S^{5}$ and $adS_{2}\times
S^{2}$} the transformation of the gravitino field strength can be brought
to
a
form which depends on the Weyl tensor and derivatives of the
form-field. In
particular it means that eq. (\ref{integra}) can be rewritten  using
Einstein's equation and one finds that it is equivalent to the vanishing
of the
Weyl tensor.  It is then
simple to observe that it is {\it the conformal flatness of these vacua}
and
the fact
that the form is constant, which force the superfields to be
supercovariant.
This was
the argument used in \cite{K, FK} with respect to Bertotti-Robinson vacuum
and
for the analysis of $R^4$ terms in \cite{banksgreen}. Now however we see
that
this is only a part of a larger picture: {\it in  odd dimension where
there are
both electric as well as magnetic supersymmetric vacua which are dual to
each other , the
metric of
$adS_{p+2}\times
S^{d-p-2}$ is not conformally flat} \cite{conf}. Still the integrability
condition for the existence of the {\it maximal unbroken supersymmetry} as
shown e. g. in M-theory case in eq. (\ref{2}) provides the
crucial vanishing of the
component of the  basic superfield depending on the curvature.

Given the strong argument for the exactness  of both the maximally
supersymmetric
flat superspace $SO(1,d-1)$-symmetric vacuum and the compactified ones
with
$SO(1, p+1)\times SO(d-p-3)$
symmetry, it is tempting to speculate that the branes which according to
\cite{GT} interpolate between
these vacua  may also be proven to be exact. This
however
may be more difficult to establish since only 1/2 of unbroken
supersymmetry is
available. The second half of supersymmetries which are broken generate
ultrashort multiplets, and all relevant superfields are
not covariantly constant but ultrashort (depend on half of $\theta$'s).
Recently an absence of corrections  from  $R^4$ terms to equations for
Reissner-Nordstrom black holes in  N=2 d=4 supergravity without matter
multiplets was demonstrated in \cite{Kelly}, using
the relation between the $W_{\alpha\beta}$-superfield of Poincar\'{e}
supergravity and the unconstrained superfield $V$ of conformal
supergravity.

 \vskip 2 cm
We had  fruitful discussion of various topics raised in this paper with T.
Banks,
S. Gubser, J. Kumar,  L.
Susskind and A. Tseytlin.
R.K. would like to acknowledge stimulating discussions of supersymmetric
$adS_{p+2}\times
S^{d-p-2}$ geometries  in the context of the superspace with P.
Howe,
D. Sorokin,  E. Sezgin, K. Stelle, M. Tonin, P. Townsend,  A. Van Proeyen,
 P.
West and B. de Wit.
This work  is supported by the NSF grant PHY-9219345.  The work of
A. R. is
also
supported in part by the Department of Energy under contract No.
DE-AC03-76SF00515.


\begin{thebibliography}{99}

\bibitem{juan} J. Maldacena, {\it The Large N limit of superconformal
field
theories and supergravity}, hep-th/9711200.
\bibitem{polyakov}S.S. Gubser, I.R. Klebanov and A.M. Polyakov,
{\it Gauge Theory Correlators from Non--critical String Theory},
hep--th/9802109.


\bibitem{witten} E. Witten, {\it Anti-de Sitter space and holography},
hep-th/9802150.

\bibitem{banksgreen} T. Banks and M. B. Green, {\it Nonperturbative
Effects
in $AdS_5\times S^5$
                  String Theory and d = 4 SUSY Yang-Mills},
hep-th/9804170.

\bibitem{MSW} J. Maldacena, A. Strominger and E. Witten, {\it Black hole
entropy in
M theory},
hep-th/9711053;  C. Vafa, {\it Black holes and Calabi-Yau threefolds},
hep-th/9711067.



\bibitem{Klaus}  K. Behrndt, G. Cardoso, B. de Wit, D. Lust, T.Mohaupt and
W. Sabra,
 {\it Higher order black hole solutions in N=2 supergravity and
                  Calabi-Yau string backgrounds}, hep-th/9801081.


\bibitem{Br1}
H. W. Brinkmann, Proc. Natl. Acad. Sci. U. S. {\bf  9} (1923) 1;
Math. Annal.
{\bf 94}, (1925) 119.


\bibitem{Gu1}
R. G\"{u}ven, Phys. Lett. {\bf 191B} (1987) 265;
D. Amati and  C. Klim\v{c}ik, Phys. Lett {\bf 219B} (1988) 443;
G.T.
Horowitz, in: {\it Proceedings of Strings '90}, College Station,
Texas, March
1990 (World Scientific, 1991) and references therein.


\bibitem{gates} S. James Gates Jr. , {\it Another Solution for N=2
Superspace
Bianchi Identities}, Phys. Lett.{\bf 96B} (1980) 305.




\bibitem{K} R. Kallosh, {\it Supersymmetric black holes}, Phys. Lett. {\bf
B282} (1992) 80,
hep-th/9201029. R. Kallosh and A. Peet, {\it Dilaton Black Holes near
Horizon},
Phys. Rev. {\bf D46} (1992)
5223; hep-th/9209116.


\bibitem{FK} S. Ferrara and R. Kallosh, {\it Supersymmetry and
attractors},
Phys. Rev.{\bf
D54} (1996) 1514, hep-th/9602136.\\
R. Kallosh, J. Rahmfeld and W.K. Wong, Phys. Rev.{\bf
D57} (1998) 1063.

\bibitem{HoweWest} P. S. Howe and P. C. West, {\it The Complete N=2, d =
10
Supergravity},
Nucl. Phys. {\bf B238} (1984) 181.

\bibitem{CremFer} E. Cremmer and S. Ferrara, {\it Formulation of
Eleven-Dimensional
Supergravity
in Superspace}, Phys. Lett.{\bf 91B} (1980) 61.

\bibitem{BrinkHowe} L. Brink and P. Howe, {\it Eleven-dimensional
Supergravity On the Mass Shell in Superspace}, Phys. Lett.{\bf 91B} (1980)
384.


\bibitem{KK} R. Kallosh and J. Kumar, {\it Supersymmetry enhancement
 of $D$-$p$-branes and $M$-branes}, Phys. Rev. {\bf D56} (1997) 4934,
hep-th/9704189.



\bibitem{TseyMets} R.R. Metsaev and A.A. Tseytlin, {\it  Type IIB
superstring
action in $AdS_5 x S^5$ background}, hep-th/9805028.




\bibitem{GT} G. W. Gibbons and P. K. Townsend, {\it Vacuum
interpolation in supergravity via super p-branes}, Phys.
Rev. Lett. {\bf 71} (1993) 3754 ; hep-th/9307049.



\bibitem{conf} P.Claus, R.Kallosh, J.Kumar, P.K.Townsend and
A.Van Proeyen, {\it Conformal theory of M2, D3, M5
and `D1+D5' branes},  hep-th/9801206.

\bibitem {Kelly} E. Cremmer, H. Lu, C.N. Pope,  K.S. Stelle, {\it Spectrum
Generating Symmetries for BPS Solitons}, hep-th/9707207.




\end{thebibliography}
\end{document}